\documentclass[prd,onecolumn,amsfonts,amsmath,amssymb,superscriptaddress]{revtex4-2}

\usepackage{adjustbox}
\usepackage{float,hyperref}
\usepackage{tabularx}
\usepackage[T1]{fontenc}
\usepackage{graphicx}
\usepackage{hhline}
\usepackage[utf8]{inputenc}
\usepackage{multirow}
\usepackage{txfonts}
\usepackage{soul}
\usepackage[svgnames]{xcolor}

\usepackage[paperheight=29.7cm,paperwidth=21.0cm,left=2.54cm,right=2.54cm,top=2.25cm,bottom=2.4cm]{geometry}
\usepackage{xurl}

\begin{document}
\title{Copenhagen Survey on Black Holes and Fundamental Physics}

\author{Alice Y. Chen}
\email{ay7chen@uwaterloo.ca}
\affiliation{Department of Physics and Astronomy, University of Waterloo, 200 University Ave W, N2L 3G1, Waterloo, Canada}
\affiliation{Waterloo Centre for Astrophysics, University of Waterloo, Waterloo, ON, N2L 3G1, Canada}
\affiliation{Perimeter Institute For Theoretical Physics, 31 Caroline St N, Waterloo, Canada}

\author{Phil Halper}
\email{philhalper1@gmail.com}
\affiliation{Independent Scholar, London, UK}

% \affiliation{Department of Physics and Astronomy, University of Waterloo, 200 University Ave W, N2L 3G1, Waterloo, Canada}
% \affiliation{Waterloo Centre for Astrophysics, University of Waterloo, Waterloo, ON, N2L 3G1, Canada}
% \affiliation{Perimeter Institute For Theoretical Physics, 31 Caroline St N, Waterloo, ON, N2L 2Y5, Canada}

\author{Niayesh Afshordi}
\email{nafshordi@pitp.ca}
\affiliation{Department of Physics and Astronomy, University of Waterloo, 200 University Ave W, N2L 3G1, Waterloo, Canada}
\affiliation{Waterloo Centre for Astrophysics, University of Waterloo, Waterloo, ON, N2L 3G1, Canada}
\affiliation{Perimeter Institute For Theoretical Physics, 31 Caroline St N, Waterloo, Canada}
\begin{abstract}
The purpose of this survey is to take a snapshot of the attitudes of physicists working on some of the most pressing questions in modern physics, which may be useful to sociologists and historians of science.  For this study, a total of 85 completed surveys were returned out of 151 registered participants of the ``Black holes Inside and out'' conference, held in Copenhagen in 2024. The survey asked questions about  some of the most contentious issues in fundamental physics, including the nature of black holes and dark energy. A number of surprising results were found.  For example, some of the leading frameworks, such as the cosmological constant, cosmic inflation, or string theory - while most popular - gain less than the majority of votes from the participants. The only statement that gains majority approval (by 68\% of participants) was that the Big Bang meant ``the universe evolved from a hot dense state'', not ``an absolute beginning time''. These results provide reasons for caution in describing ideas as consensus in the scientific community when a more nuanced view may be justified.
\end{abstract}

\maketitle
\section{Introduction}

In August 2024, \textit{Black Holes Inside and Out} conference was held with a multi-disciplinary approach to black hole physics.  The conference was organized by the Niels Bohr Institute and held at the iconic Black Diamond Building in Copenhagen, Denmark. It included leading theorists and experimentalists in astrophysics, general relativity, numerical relativity, gravitational wave astronomy, and cosmology. We considered this a good opportunity to survey experts on current controversies within physics.  Sociologists and historians of science could find these results useful for studying the scientific process in controversial areas of physics, circa 2024.  We also compare the results of this survey to popular reports of what physicists believe and highlight some significant discrepancies. The paper is organized as follows: we start by a brief description of our methodology in Section \ref{sec:methods}, Section \ref{sec:blackholes} focuses on topics within black hole physics, while Section \ref{sec:fundamental} focuses on broader issues within fundamental physics. The last section summarizes the highlights and concludes the paper.

\section{Methods}\label{sec:methods}

A list of questions was proposed by the authors and then approved by conference organisers. The survey was announced during several sessions and distributed in the lobby. A total of 85 completed surveys were returned out of 151 participants listed as attendants on the conference web site.  However, for some questions, a few respondents ticked more than one answer (see Section \ref{sec:Appendix}). Only questions that had single answers from a participant  were included in our main results. In the appendix, we show the addition of these multiple votes makes little difference to the overall picture.  Results were compiled by a third party who was not involved in the survey design. 

\section{Black Holes}\label{sec:blackholes}

In this section, we report on the results of controversial topics within the area of black hole physics. 

\subsection{Information Paradox}

\begin{figure}
    \centering    \includegraphics[width=0.75\linewidth]{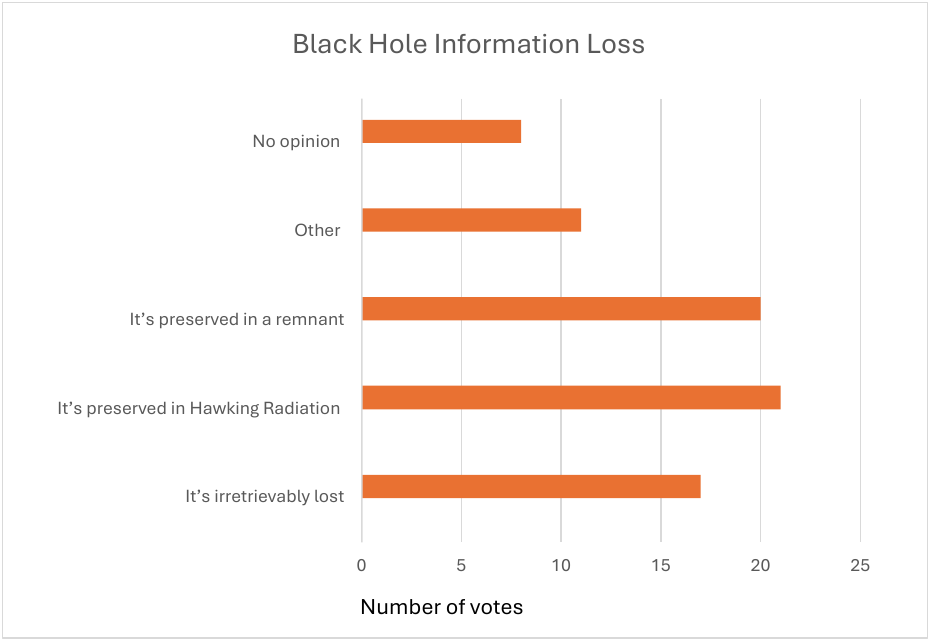}
    \caption{\small In your opinion, what do you think is the most likely correct description of the information that falls into a black hole?}
    \label{fig:Q1}
\end{figure}

Since the 1970’s, it has been argued that black holes in general relativity destroy information, potentially violating the unitarity principle of quantum mechanics  \cite{hawking1976breakdown}. Several solutions to this paradox have been proposed  \cite{mathur2009the} but no consensus currently exists. The results gathered from this survey show that most physicists (53$\%$) do believe that information is preserved either in a remnant (26$\%$), or in Hawking radiation (27$\%$). However, there is a significant minority (22$\%$) that believes information is permanently lost in black holes, which is in tension with the claim amongst experts that the issue is now settled by the holographic principle \cite{susskind2008the}. In fact, both remnants and the destruction of information are in conflict with the holographic principle, as remnants have more entropy than their area permits, and information should always be preserved on the holographic boundary. The proponents of these latter two ideas add up to nearly half of the respondents (see Figure \ref{fig:Q1} and Table \ref{tab:Q1}).

% \textcolor[HTML]{222222}{}
% \vspace{1\baselineskip}
% \textcolor{red}{Graphics Type `CHART' is not supported yet. Please insert it as image.}\par

\begin{table}[H]
    \centering
    \begin{tabular}
     {|c|c|c|}
     \hline 
     \textbf{Answer Option} & Number of Votes & Percentage \\ 
     \hline
     It’s irretrievably lost & 17 & 22$\%$ \\
     \hline
     It’s preserved in Hawking Radiation  & 21 & 27$\%$ \\ 
     \hline 
     It’s preserved in a remnant & 20 & 26$\%$ \\ 
     \hline
     Other & 11 & 14$\%$ \\ 
     \hline
     No opinion & 8 & 10$\%$ \\
     \hline 
     Total & 77 & $100\%$  \\
    \hline
    \end{tabular}
    \caption{\small In your opinion, what do you think is the most likely correct description of the information that falls into a black hole?}
    \label{tab:Q1}
\end{table}

\vspace{16\baselineskip}
\subsection{Supermassive Black Hole Formation}

\begin{figure}[H]
    \centering    \includegraphics[width=0.75\linewidth]{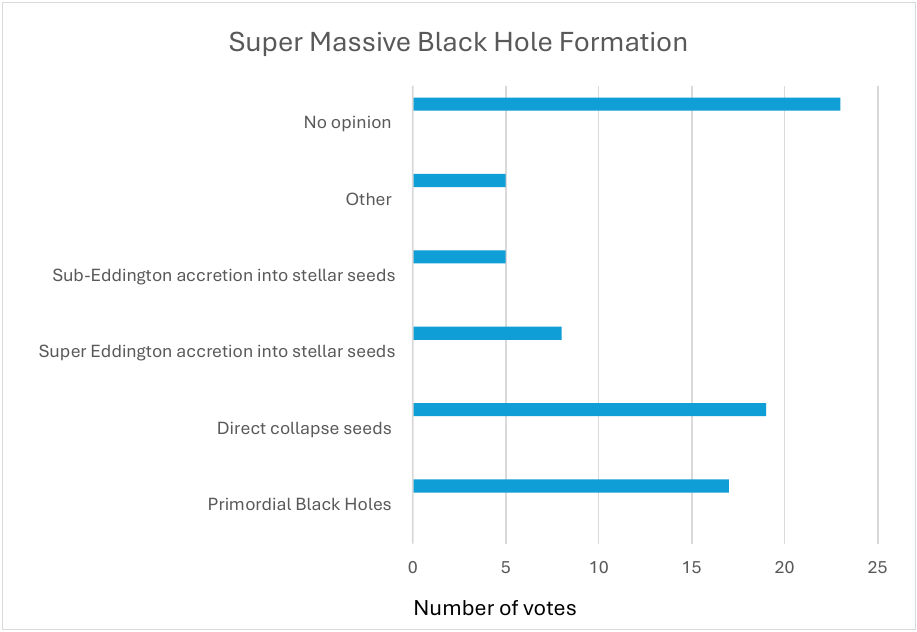}
    \caption{\small In your opinion, what is the most likely scenario for the creation of supermassive black holes?}
    \label{fig:Q2}
\end{figure}

Supermassive black holes have been measured with masses larger than $4\times 10^{10} M_{\odot}$ \cite{shemmer2004near} but their formation process remains a mystery. Our results show that out of those that expressed an opinion, direct collapse (25$\%$) was the most popular option with primordial black holes (22$\%$) being a close second (see Figure \ref{fig:Q2} and Table \ref{tab:Q2}).

\begin{table}[H]
    \centering
    \begin{tabular}
    {|c|c|c|}
    \hline 
    \textbf{Answer Option} & Number of Votes & Percentage \\ 
    \hline
    Primordial Black Holes & 17 & 22$\%$ \\
    \hline 
    Direct collapse seeds & 19 & 25$\%$ \\ 
    \hline
    Super Eddington accretion into stellar seeds & 8 & 10$\%$ \\
    \hline 
    Sub-Eddington accretion into stellar seeds & 5 & 6$\%$ \\ 
    \hline 
    Other & 5 & 6$\%$ \\ 
    \hline 
    No opinion & 23 & 30$\%$ \\ 
    \hline
    Total & 77 & $100\%$ \\
    \hline
    \end{tabular}
    \caption{\small In your opinion, what is the most likely scenario for the creation of supermassive black holes?}
    \label{tab:Q2}
\end{table}

%\vspace{12\baselineskip}
\subsection{Fate of Matter in Black Holes}

\begin{figure}[H]
    \centering    \includegraphics[width=0.75\linewidth]{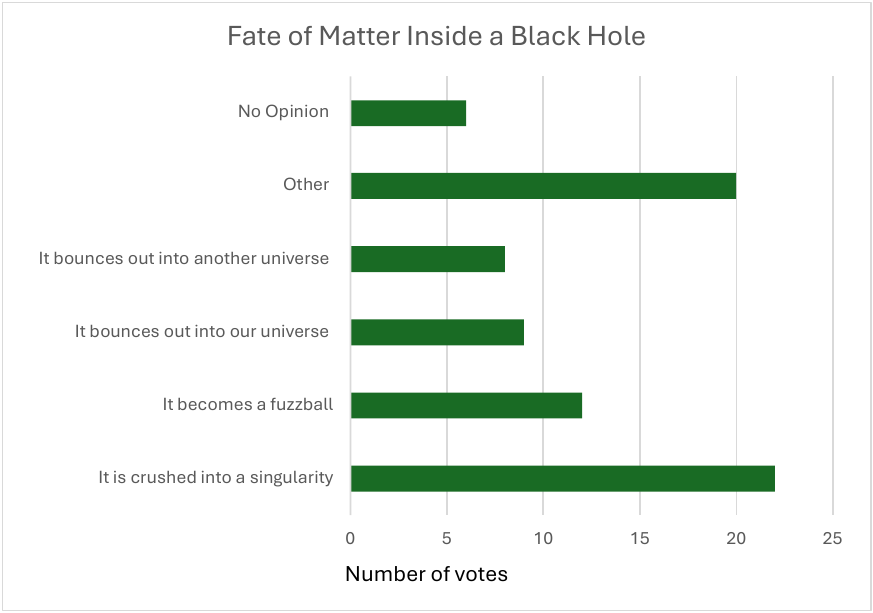}
    \caption{\small In your opinion, what is likely to happen to the matter that falls into a black hole?}
    \label{fig:Q3}
\end{figure}

In classical relativity, it has been proven that matter falling into a black hole will be crushed into a singularity \cite{penrose1965gravitational}. But do physicists believe this really happens, or are they happy to entertain more exotic alternatives? The classical view remains favoured by a substantial portion of respondents (29$\%$, see Figure \ref{fig:Q3} and Table \ref{tab:Q3}). While no individual alternative to the singularity problem garners a higher level of approval, the combined backing for the three proposed exotic solutions (38$\%$) exceeds that of the classical option.

\begin{table}[H]
    \centering
    \begin{tabular}
    {|c|c|c|}
    \hline 
    \textbf{Answer Option} & Number of Votes & Percentage \\ 
    \hline
    It is crushed into a singularity & 22 & 29$\%$ \\ 
    \hline 
    It becomes a fuzzball & 12 & 16$\%$ \\ 
    \hline 
    It bounces out into our universe & 9 & 12$\%$ \\ 
    \hline 
    It bounces out into another universe & 8 & 10$\%$ \\ 
    \hline 
    Other & 20 & 26$\%$ \\ 
    \hline 
    No Opinion & 6 & 8$\%$ \\ 
    \hline 
    Total & 77 & $100\%$ \\
    \hline
    \end{tabular}
    \caption{\small In your opinion, what is likely to happen to the matter that falls into a black hole?}
    \label{tab:Q3}
\end{table}

%\vspace{24\baselineskip}
\subsection{Testing Quantum Gravity Models with Black Holes }

\begin{figure}[H]
    \centering    \includegraphics[width=0.73\linewidth]{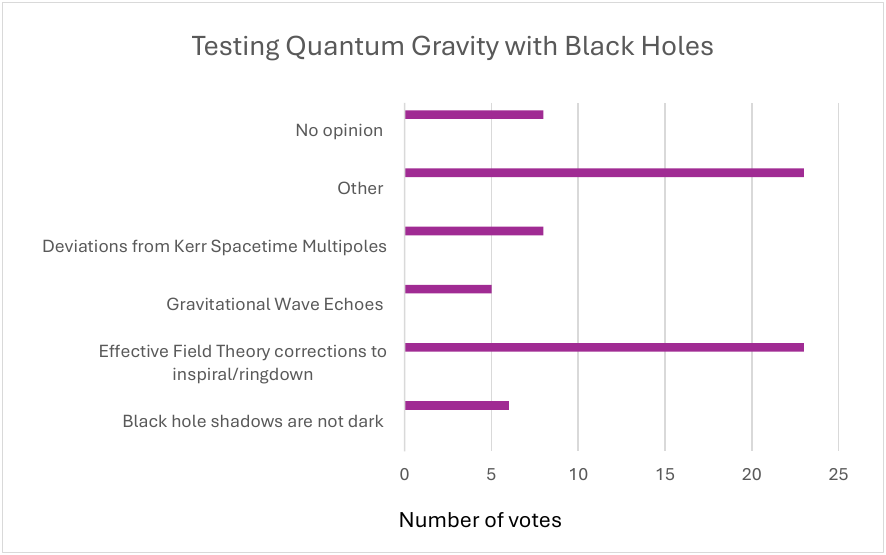}
    \caption{\small In your opinion, what are the most promising observable signature of quantum gravity in black holes?}
    \label{fig:Q4}
\end{figure}

The energy scale of quantum gravity is far beyond the reach of any particle detector on Earth. However, it has been shown that a black hole can accelerate particles to arbitrarily high levels \cite{bañados2009kerr}. This provides hope that black holes could be used as probes of quantum gravity. However, there is more than one possible way to use black holes for this purpose. In our survey, we see that among those who selected a particular method, Effective Field Theory corrections to inspiral/ringdown was by far the most popular view (32$\%$, see Figure \ref{fig:Q4} and Table \ref{tab:Q4}). However, the same number of respondents suggested that there could be other methods to test quantum gravity with black holes, not listed in our survey. The nature of such methods remains an enigma to the authors of this article. 

\begin{table}[H]
    \centering
    \begin{tabular}{|c|c|c|}
    \hline 
    \textbf{Answer Option} & Number of Votes & Percentage \\
    \hline 
    No opinion & 8 & 11$\%$ \\
    \hline 
    Other & 23 & 32$\%$ \\
    \hline
    Deviations from Kerr Spacetime Multipoles & 8 & 11$\%$ \\
    \hline 
    Gravitational Wave Echoes & 5 & 7$\%$ \\
    \hline 
    Effective Field Theory corrections to inspiral/ringdown & 23 & 32$\%$ \\
    \hline 
    Black hole shadows are not dark & 6 & 8$\%$ \\
    \hline 
    Total & 73 & $100\%$  \\
    \hline
    \end{tabular}
    \caption{\small In your opinion, what are the most promising observable signature of quantum gravity in black holes?}
    \label{tab:Q4}
\end{table}

%\vspace{47\baselineskip}
\section{Fundamental Physics }\label{sec:fundamental}

In this section we report on results from questions concerning a broader range of controversies within physics not specifically linked to black holes. 

\subsection{Interpretation of Quantum Mechanics}

\begin{figure}[H]
    \centering    \includegraphics[width=0.84\linewidth]{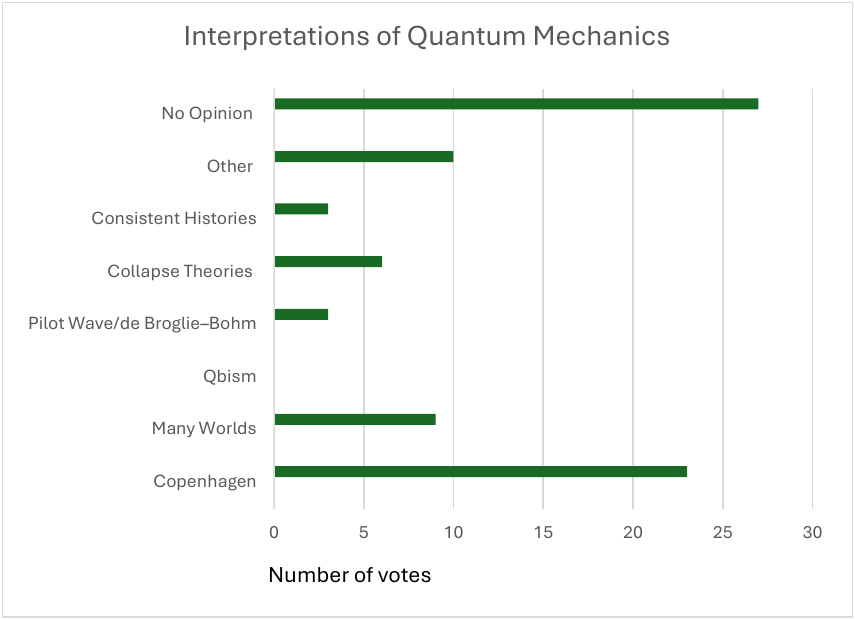}
    \caption{\small In your opinion, which interpretation of quantum mechanics is most likely to be correct?}
    \label{fig:Q5}
\end{figure}

Quantum mechanics has been described as both a great success in physics but also ``a scandal"  \cite{wallace2016philosophy} in that despite a remarkable predictive accuracy, its physical interpretation remains elusive. Our survey shows the two most well-known ``interpretations" of quantum mechanics, Copenhagen and Many Worlds, are indeed the most popular choices amongst those who chose a specific option, with Copenhagen remaining the preferred option (see Figure \ref{fig:Q5} and Table \ref{tab:Q5}). Copenhagen’s dominance is consistent with previous surveys \cite{schlosshauer2013a, tegmark1998the}).

\begin{table}[H]
    \centering
    \begin{tabular}{|c|c|c|}
    \hline 
    \textbf{Answer Option} & Number of Votes & Percentage \\
    \hline
    Copenhagen & 23 & 28$\%$ \\
    \hline 
    Many Worlds & 9 & 11$\%$ \\
    \hline
    Qbism & 0 & 0$\%$ \\
    \hline 
    Pilot Wave/de Broglie-Bohm & 3\ & 4$\%$ \\
    \hline 
    Collapse Theories & 6 & 7$\%$ \\
    \hline 
    Consistent Histories & 3 & 4$\%$ \\
    \hline
    Other & 10 & 12$\%$ \\
    \hline 
    No Opinion & 27 & 33$\%$ \\
    \hline 
    Total & 81 & $100\%$  \\
    \hline
    \end{tabular}
    \caption{\small In your opinion, which interpretation of quantum mechanics is most likely to be correct?}
    \label{tab:Q5}
\end{table}

%\vspace{3\baselineskip}
\subsection{Quantum Gravity}

\begin{figure}[H]
    \centering    \includegraphics[width=0.72\linewidth]{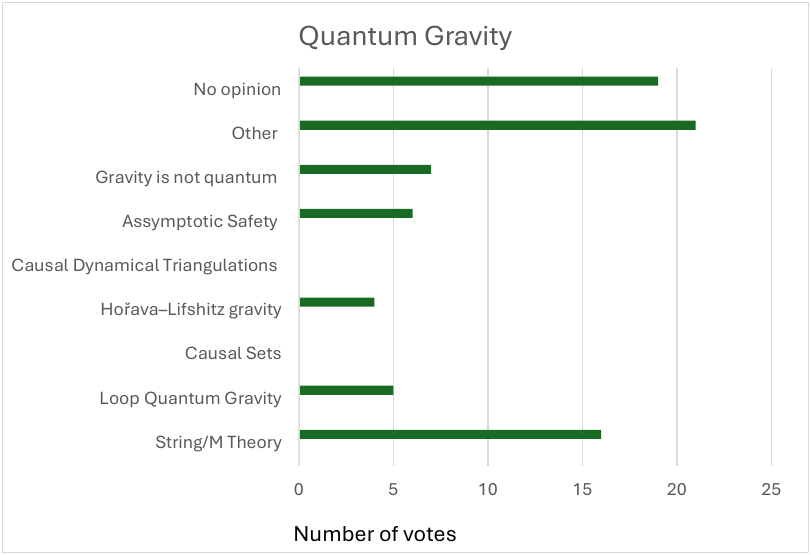}
    \caption{\small In your opinion, which the best candidate for a theory of quantum gravity?}
    \label{fig:Q6}
\end{figure}

A theory of quantum gravity has been described as the ``holy grail" of physics \cite{rovelli2006unfinished} but no candidate has been widely accepted by the scientific community. String theory, perhaps the most prominent proposal, has been described by commentators as anything from ``the only idea about quantum gravity with any substance'' \cite{armas_2021} to a theory that is ``dead''  \cite{Papazoglou2023}. Participants in our survey who expressed a preference overwhelmingly favor string theory over other candidates. However, consensus remains elusive, with some researchers even questioning whether gravity is quantum — an idea recently explored by \cite{oppenheim2023a}. Another surprising result is that asymptotically safe quantum gravity was the second most popular theory, even though a google scholar search for the third most popular candidate (loop quantum gravity) returned more than ten times the results than asymptotic safety. Interestingly, this is the only question where those picking none of the presented physical options (combining no opinion and other) represented the majority view (51$\%$, see Figure \ref{fig:Q6} and Table \ref{tab:Q6}). 

\begin{table}[H]
    \centering
    \begin{tabular}{|c|c|c|}
    \hline 
    \textbf{Answer Option} & Number of Votes & Percentage \\
    \hline
    String/M Theory & 16 & 21$\%$ \\ 
    \hline
    Loop Quantum Gravity & 5 & 6$\%$ \\ 
    \hline
    Causal Sets & 0 & 0$\%$ \\ 
    \hline
    Hořava–Lifshitz gravity & 4 & 5$\%$ \\ 
    \hline
    Causal Dynamical Triangulations & 0 & 0$\%$ \\ 
    \hline
    Asymptotic Safety & 6 & 8$\%$ \\ 
    \hline
    Gravity is not quantum & 7 & 9$\%$ \\ 
    \hline
    Other & 21 & 27$\%$ \\ 
    \hline 
    No opinion & 19 & 24$\%$ \\ 
    \hline
    Total & 78 & 100$\%$ \\
    \hline 
  \end{tabular}
  \caption{\small In your opinion, which the best candidate for a theory of quantum gravity?}
  \label{tab:Q6}
\end{table}

%\vspace{11\baselineskip}
\subsection{Dark Matter}

\begin{figure}[H]
    \centering    \includegraphics[width=0.75\linewidth]{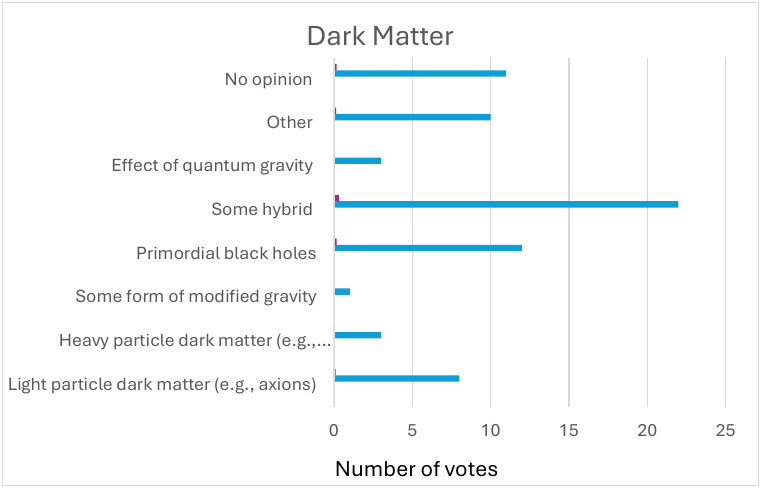}
    \caption{\small In your opinion, what is likely to be the predominant explanation for the apparent dark matter in the universe?}
    \label{fig:Q7}
\end{figure}

The standard model of cosmology, LCDM, assumes most of the matter in the universe is ``dark'', meaning that it does not interact with electromagnetic radiation \cite{turner2018λ}. However, establishing which of the many dark matter candidates is part of nature (and which are not) has not been accomplished by science. According to Wikipedia, ``The most prevalent explanation is that dark matter is some as-yet-undiscovered subatomic particle'' \cite{DM_wiki}. However, in our survey, the most popular specific candidate is ``primordial black holes'', which had more votes than WIMPs and axions combined (although in the appendix we see when multiple votes are counted, WIMPS do have a significant boost). Nevertheless, the fact that this survey was performed at a black hole conference may be a contributing factor to this result. Modified gravity theories remain on the fringe, although presumably the hybrid option could include some form of modification to gravity as has been suggested (e.g., in the superfluid model \cite{berezhiani2015theory}).  The hybrid explanation is the most popular (31$\%$, see Figure \ref{fig:Q7} and Table \ref{tab:Q7}) of all options. A hybrid might also refer to a combination of heavy and light particles, primordial black holes, and some modification to gravity. Nonetheless, 76\% of the respondents with an opinion on the matter appear to believe in some type of dark matter. 

\begin{table}[]
    \centering
    \begin{tabular}{|c|c|c|}
    \hline
    \textbf{Answer Option} & Number of Votes & Percentage \\
    \hline
    Light particle dark matter (e.g., axions & 8 & $11\%$ \\
    \hline
    Heavy particle dark matter (e.g., WIMPs) & 3 & $4\%$ \\
    \hline
    Some form of modified gravity & 1 & $1\%$ \\
    \hline 
    Primordial black holes & 12 & $17\%$ \\
    \hline 
    Some hybrid of the above & 22 & $31\%$ \\
    \hline 
    Effect of quantum gravity & 3 & 4$\%$ \\
    \hline 
    Other & 10 & $14\%$ \\
    \hline 
    No opinion & 11 & $16\%$ \\
    \hline 
    Total & 70 & 100$\%$ \\
    \hline 
    \end{tabular}
     \caption{\small In your opinion, what is likely to be the predominant explanation for the apparent dark matter in the universe?}
     \label{tab:Q7}
\end{table}

%\vspace{10\baselineskip}
\subsection{Early Universe Cosmology }

\begin{figure}[H]
    \centering    \includegraphics[width=0.75\linewidth]{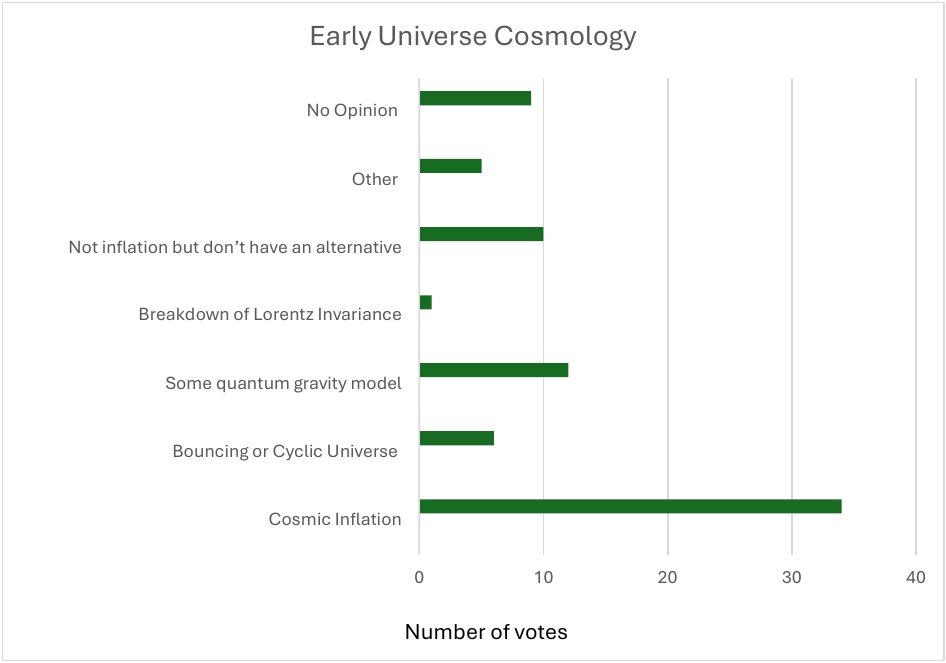}
    \caption{\small In your opinion, issues such as the horizon problem, flatness, and origin of structure problem are most likely solved by which scheme?}
    \label{fig:Q8}
\end{figure}

The standard picture of the Big Bang faces several difficult puzzles such as the horizon, monopole, flatness, and the origin of structure problem \cite{guth2007eternal}. Inflation offers solutions to these challenges and has been described as the third pillar of the standard LCDM model \cite{turner2018λ}. However, our survey reveals that while it remains the most popular choice, it - like string theory —does not command majority support. That said, its 44\% backing (see Figure \ref{fig:Q8} and Table \ref{tab:Q8}) brings it significantly closer to consensus compared to the 21\% support for string theory among quantum gravity proposals. 

\begin{table}[H]
    \centering
    \begin{tabular}{|c|c|c|}
    \hline
    \textbf{Answer Option} & Number of Votes & Percentage \\
    \hline
    Cosmic Inflation & 34 &  44$\%$ \\ 
    \hline 
    Bouncing or Cyclic Universe &  6 & 8$\%$ \\
    \hline 
    Some quantum gravity model & 12 &  16$\%$ \\ 
    \hline 
    Breakdown of Lorentz Invariance & 1 &  1$\%$ \\ 
    \hline 
    Not inflation but don’t have an alternative &  10 & 13$\%$ \\
    \hline 
    Other  &  5 &  6$\%$ \\ 
    \hline 
    No Opinion  &  9 &  12$\%$ \\ 
    \hline 
    Total & 77 & 100$\%$ \\
    \hline 
    \end{tabular}
    \caption{\small In your opinion, issues such as the horizon problem, flatness, and origin of structure problem are most likely solved by which scheme?}
    \label{tab:Q8}
\end{table}

%\vspace{10\baselineskip}
\subsection{Meaning of the Big Bang}

\begin{figure}[H]
    \centering    \includegraphics[width=0.75\linewidth]{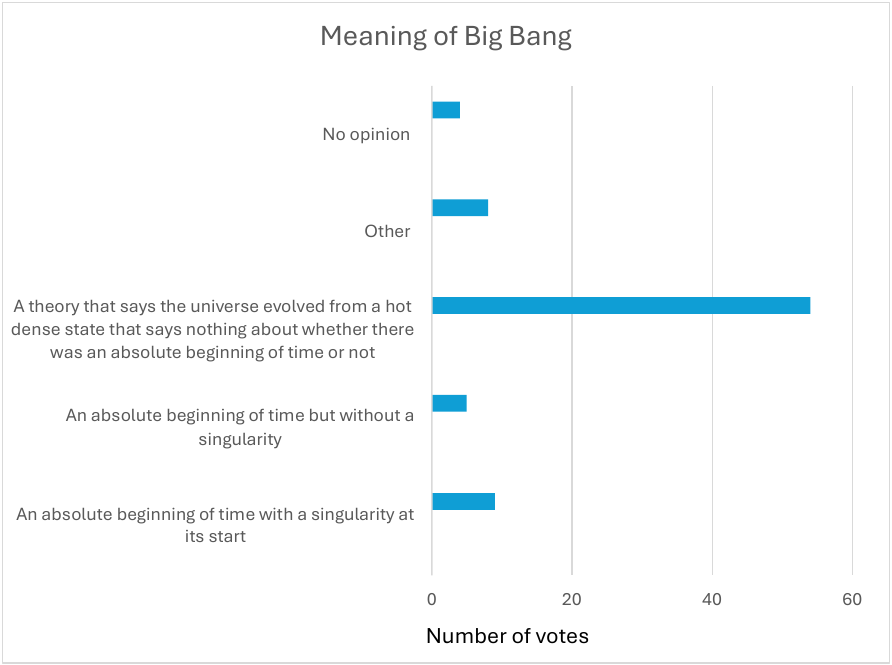}
    \caption{\small In your opinion, how should we understand the Big Bang?}
    \label{fig:Q9}
\end{figure}

It has been claimed that the Big Bang represents the beginning of time and this is a view now ``taken for granted" \cite{ hawking2012the}. However, a challenge comes from Guth, who claimed that the Big Bang is ``a theory of the aftermath of the bang" that says, ``nothing about what banged, why it banged, or what happened before it banged"  \cite{bradt20143}. The very meaning of the term Big Bang is thus not agreed upon. Our survey results strongly favour Guth’s view over Hawking’s. The Big Bang making no claim about whether time had a beginning or not and should be understood only as the theory that says the universe evolved from a hot dense state enjoys a rare majority support of 68$\%$. In that sense, it is perceived as a theory of what happened after the Bang only. Notably, there is more consensus on this issue than any other issue we surveyed. However, a counter-argument to this conclusion may be that for this question, there were fewer options. The classical singularity is taken far less seriously (11$\%$) than in the case of black holes (29$\%$). The number of No Opinions on this topic was also the fewest (5$\%$) of any topic surveyed here (see Figure \ref{fig:Q9} and Table \ref{tab:Q9}). An important implication of this result is that it may not be correct to say that scientists believe that the universe has a beginning, or even that it is 13.8 billion years old, rather it is at least 13.8 billion years old. 

\begin{table}[H]
    \centering
    \begin{tabular}{|p{0.52\linewidth}| p{0.1\linewidth}|p{0.1\linewidth}|}
    \hline
    \textbf{Answer Option} & Number of Votes & Percentage \\
    \hline
    An absolute beginning of time with a singularity at its start & 9 & 11$\%$ \\ 
    \hline 
    An absolute beginning of time but without a singularity & 5 &  6$\%$ \\ 
    \hline 
    A theory that says the universe evolved from a hot dense state that says nothing about whether there was an absolute beginning of time or not &  54 & 68$\%$ \\
    \hline 
    Other & 8 & 10$\%$ \\
    \hline 
    No opinion & 4 & 5$\%$ \\
    \hline 
    Total & 80 & $100\%$ \\
    \hline 
    \end{tabular}
    \caption{\small In your opinion, how should we understand the Big Bang?}
    \label{tab:Q9}
\end{table}

%\vspace{1\baselineskip}
\subsection{Anthropic Coincidences and Fine tuning}

\begin{figure}[H]
    \centering    \includegraphics[width=0.73\linewidth]{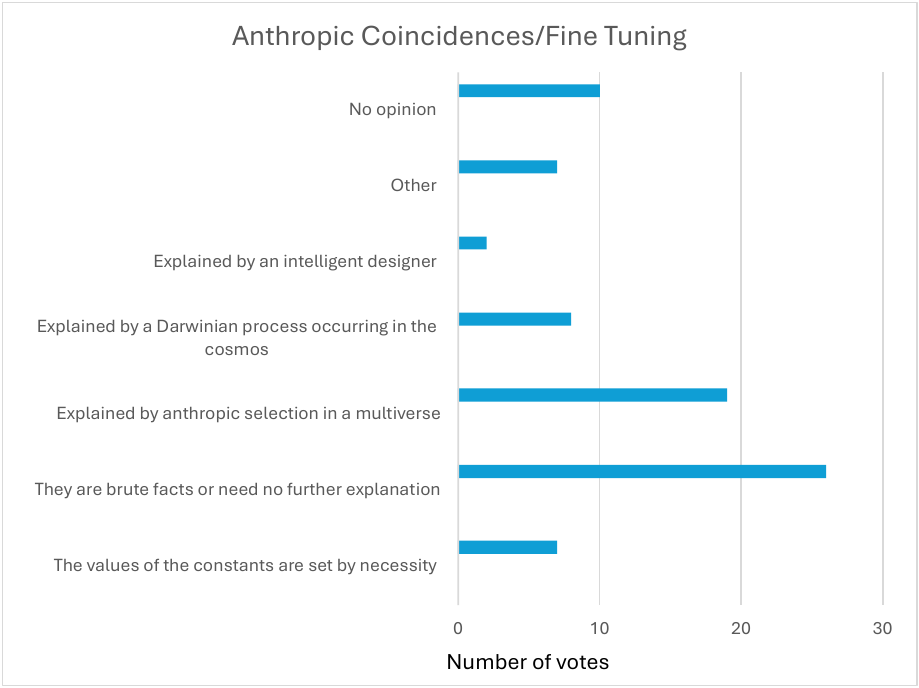}
    \caption{\small In your opinion, what explains the values of physical constants of nature and the claimed anthropic coincidences?}
    \label{fig:Q10}
\end{figure}

The constants of nature are free parameters in current physical theories. But what explains their values? Some have claimed that they are ``fine-tuned for life'', which requires exotic explanations either in the form of anthropic selection in a multiverse or an intelligent designer \cite{lewis2016a}. A similar question was asked of philosophers in the PhilPapers survey \cite{bourget2023philosophers}.  In the PhilPapers survey, the most popular choice was that the constants are considered brute facts and thus require no exotic explanation. For philosophers, that number was 32$\%$; for our survey of physicists, this was also the most popular choice and the number was remarkably similar at 33$\%$. However in our case, anthropic selection in a multiverse was a clear second (24$\%$, see Figure \ref{fig:Q10} and Table \ref{tab:Q10}) whereas for philosophers this option was fourth. One possibility that was not considered in the PhilPapers survey was a Darwinian solution to the problem (see \cite{smolin1997the} for an example). Here, that option was the third most popular at 10$\%$.

\begin{table}[H]
    \centering
    \begin{tabular}{|p{0.53\linewidth}| p{0.1\linewidth}|p{0.1\linewidth}|}
    \hline
    \textbf{Answer Option} & Number of Votes & Percentage \\
    \hline
    The values of the constants are set by necessity & 7 & $9\%$ \\
    \hline 
    They are brute facts or need no further explanation & 26 & $33\%$ \\
    \hline 
    Explained by anthropic selection in a multiverse & 19 & $24\%$ \\
    \hline 
    Explained by a Darwinian process occurring in the cosmos & 8 & $10\%$ \\
    \hline 
    Explained by an intelligent designer & 2 & $3\%$ \\
    \hline 
    Other & 7 & $9\%$ \\
    \hline 
    No opinion & 10 & $13\%$ \\
    \hline 
    Total & 79 & $100\%$ \\	
    \hline 
    \end{tabular}
    \caption{\small In your opinion, what explains the values of physical constants of nature and the claimed anthropic coincidences?}
    \label{tab:Q10}
\end{table}

%\multicolumn{1}{p{17.26cm}}

\subsection{Dark Energy} 
\begin{figure}[H]
    \centering    \includegraphics[width=0.73\linewidth]{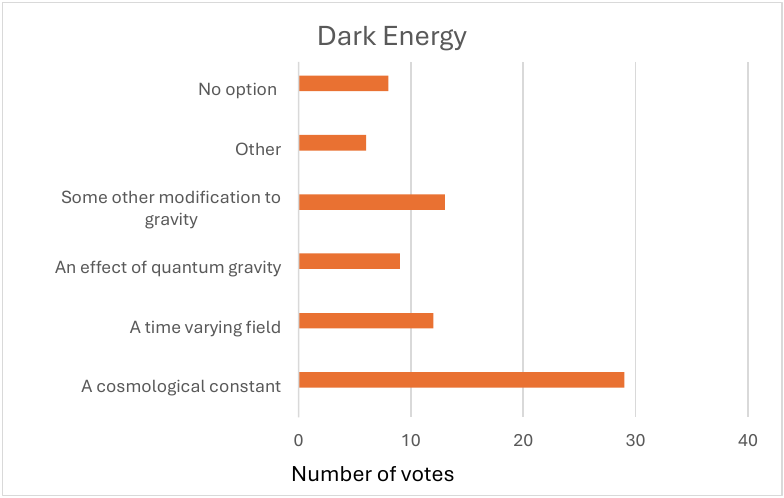}
    \caption{\small In your opinion, what is the most likely candidate to be causing the universe to accelerate in its expansion?}
    \label{fig:Q11}
\end{figure}

In 1998 it was discovered that the universe's expansion is accelerating \cite{riess1998observational}. In the standard model of cosmology, this is assumed to be driven by a cosmological constant \cite{turner2018λ}. In our survey, a cosmological constant is the most popular option (38$\%$), with significantly higher support than the next most popular option, a modification to gravity at 17$\%$ (see Figure \ref{fig:Q11} and Table \ref{tab:Q11}). However, it did not gain majority support, casting doubt that this is the consensus amongst physicists. An important implication for science communicators is that it may not be correct to say that scientists agree the ultimate fate of the universe is a ``heat death". This would be the case if there was a majority view for a cosmological constant. However, our survey shows this option falling below the threshold of support required.  

\begin{table}[H]
    \centering
    \begin{tabular}{|c|c|c|}
    \hline
    \textbf{Answer Option} & Number of Votes & Percentage \\
    \hline
    A cosmological constant & 29 & 38$\%$ \\ 
    \hline 
    A time varying field & 12 &  16$\%$ \\ 
    \hline 
    An effect of quantum gravity &  9 & 12$\%$ \\
    \hline 
    Some other modification to gravity & 13 & 17$\%$ \\ 
    \hline 
    Other & 6 & 8$\%$ \\ 
    \hline 
    No option & 8 & 10$\%$ \\ 
    \hline 
    Total & 77 & 100$\%$ \\
    \hline 
    \end{tabular}
    \caption{\small In your opinion, what is the most likely candidate to be causing the universe to accelerate in its expansion?}
    \label{tab:Q11}
\end{table}

%\vspace{9\baselineskip}
\subsection{Hubble tension }

\begin{figure}[H]
    \centering    \includegraphics[width=0.75\linewidth]{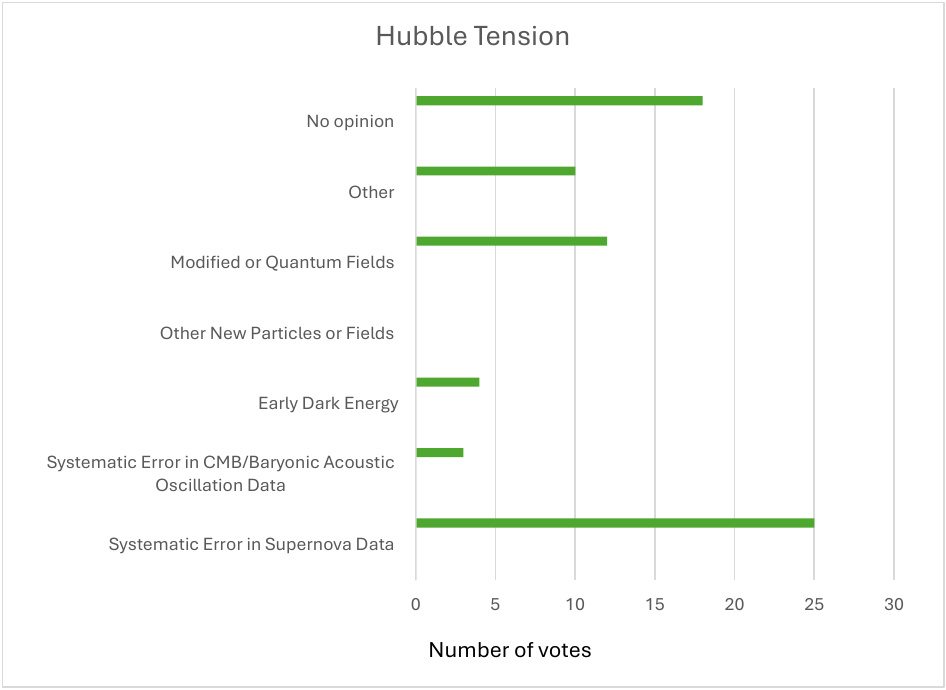}
    \caption{\small In your opinion, what is the most likely expansion for the Hubble tension?}
    \label{fig:Q12}
\end{figure}

Supernovae Ia and the cosmic microwave background (CMB) measurements can provide a value for the Hubble parameter and, in theory, should align, but they don’t \cite{valentino2021in}. This has been described by the press as a ``crisis in cosmology" \cite{phys_crisis}. Our survey shows that the most popular answer by far (38$\%$, see Figure \ref{fig:Q12} and Table \ref{tab:Q12}) is that the Hubble tension does not represent a sign of new physics, but rather systematic errors in supernova data. 

\begin{table}[H]
    \centering
    \begin{tabular}{|c|c|c|}
    \hline
    \textbf{Answer Option} & Number of Votes & Percentage \\
    \hline
    Systematic Error in Supernova Data & 25 & 35$\%$ \\
    \hline 
    Systematic Error in CMB/Baryonic Acoustic Oscillation Data & 3 & 4$\%$ \\
    \hline 
    Early Dark Energy & 4 & $6\%$ \\
    \hline 
    Other New Particles or Fields & 0 & $0\%$ \\
    \hline 
    Modified or Quantum Fields & 12 & $17\%$ \\
    \hline 
    Other &	10 & $14\%$ \\
    \hline 
    No opinion & 18 & 25$\%$ \\
    \hline
	Total & 72 & $100\%$ \\
    \hline 
    \end{tabular}
    \caption{\small In your opinion, what is the most likely expansion for the Hubble tension?}
    \label{tab:Q12}
\end{table}

%\vspace{5\baselineskip}
\section{Conclusion}

Polling scientists does not decide the resolution of scientific controversies. However, it can be useful for sociologists and historians of science to understand how scientific views may change over time. It is also useful for communicators of science to know what leading researchers in a field think. 

Surveys can also be a tool to combat publication bias. For example, it may be the case that within the literature, there is more discussion of new physics as a solution to the Hubble tension, as opposed to systematic errors. But this could mask wide skepticism within the field about the need for exotic proposals.  Counting citations might then be an incomplete probe to understand scientific opinion or gauge the extent of scientific consensus on a topic. 

Any survey will have the limitation of its sample size and ours is no exception. A conference on black holes might give different results compared to one on particle physics, foundation of physics, cosmology or solid-state physics.  However, \textit{The Black Holes Inside and Out} conference was multidisciplinary (although still within the realm of physics), providing some reason to think it may capture the attitudes of a wider scientific community than more specialised conferences. 

Several results reveal tensions between how science is often communicated and what the scientists in our survey believe. Perhaps the biggest discrepancy is the view that The Big Bang represents the beginning of time; most physicists in our survey see the theory making much more modest claims than that. There seems to be more uncertainty for the fate of the universe, as most scientists we surveyed are not convinced that the accelerated expansion is driven by a cosmological constant. Another notable discrepancy is for the claim that the ``fine tuning'' of the constants of nature forces us to choose between a multiverse or an intelligent designer. In our survey, brute facts that need no explanation was a more popular option than both; moreover, necessity and a Darwinian explanation were also both more popular than an intelligent designer. In other areas, several leading candidates for solving outstanding issues in physics are the most popular option amongst scientists that expressed a preference.  But in each case, they fail to garner majority ($\geq 50\%$) support. Examples include string theory as a theory of quantum gravity, inflation as a theory of the early universe, a cosmological constant as the best explanation for the accelerated expansion, black hole information preserved as Hawking radiation, matter being crushed into a singularity inside a black hole, and the Copenhagen interpretation of quantum mechanics.  Particle dark matter also seems less preferred than is often communicated, although in this case a preference for primordial black holes may be biased by the selection effect of being at a black hole physics conference.  We see no reason though to think this selection effect biased any of the other topics we surveyed.  This survey highlights the importance of considering a more nuanced perspective when describing ideas as consensus within the scientific community, as there may be valid reasons for caution.

\begin{acknowledgments}
We would like to thank the organizers of the ``Black Holes Inside Out'' conference, especially Vitor Cardoso and Julie de Molade, for making this survey possible, and helping with distribution and collection of the questionnaires.

NA and AC are funded by the University of Waterloo, the National Science and Engineering Research Council of Canada (NSERC) and the Perimeter Institute for Theoretical Physics.
Research at Perimeter Institute is supported by the Government of Canada through Industry Canada and by the Province of Ontario through the Ministry of Economic Development \& Innovation. 
\end{acknowledgments}

\section{Appendix}
\label{sec:Appendix}

Here we show results where completed surveys  with multiple votes per question were included, few significant differences are seen from the main results.  Two exceptions are for dark matter, where WIMPS gained a boost from 3 to 9 votes (4$\%$ to 10$\%$) and testing quantum gravity with gravitational wave echoes were boosted from 5 to 13 votes (7 $\%$ to 14$\%$).

\subsubsection{Black Hole Information Loss}

\begin{table}[H]
    \centering
    \begin{tabular}{|c|c|c|}
    \hline
    \textbf{Answer Option} & Number of Votes & Percentage \\
    \hline
    It’s irretrievably lost & 18 & 20$\%$ \\ 
    \hline 
    It’s preserved in Hawking Radiation & 25 & 27$\%$ \\ 
    \hline 
    It’s preserved in a remnant & 26 & 28$\%$ \\ 
    \hline 
    Other & 15 & 16$\%$ \\ 
    \hline 
    No opinion & 8 & 9$\%$ \\ 
    \hline 
    Total & 92 & 100$\%$ \\
    \hline 
    \end{tabular}
    \caption{\small Question 1 with some participants responding multiple times (see Figure \ref{fig:Q1} and Table \ref{tab:Q1} for comparison).}
\end{table}

\subsubsection{Super Massive Black Hole Formation }

\begin{table}[H]
    \centering
    \begin{tabular}{|c|c|c|}
    \hline
    \textbf{Answer Option} & Number of Votes & Percentage \\
    \hline
    Primordial Black Holes & 20 & 22$\%$ \\ 
    \hline 
    Direct collapse seeds & 23 & 25$\%$ \\ 
    \hline 
    Super Eddington accretion into stellar seeds & 13 & 14$\%$ \\ 
    \hline 
    Sub-Eddington accretion into stellar seeds & 7 & 8$\%$ \\ 
    \hline 
    Other & 5 & 5$\%$ \\ 
    \hline 
    No opinion & 23 & 25$\%$ \\ 
    \hline 
    Total & 91 & 100$\%$ \\
    \hline 
    \end{tabular}
    \caption{\small Question 2 with some participants responding multiple times (see Figure \ref{fig:Q2} and Table \ref{tab:Q2} for comparison).}
\end{table}

\subsubsection{Fate of Matter Inside a Black Hole }

\begin{table}[H]
    \centering
    \begin{tabular}{|c|c|c|}
    \hline
    \textbf{Answer Option} & Number of Votes & Percentage \\
    \hline
    It is crushed into a singularity & 25 & 27$\%$ \\ 
    \hline 
    It becomes a fuzzball & 15 & 16$\%$ \\ 
    \hline 
    It bounces out into our universe & 11 & 12$\%$ \\ 
    \hline 
    It bounces out into another universe & 12 & 13$\%$ \\ 
    \hline 
    Other & 22 & 24$\%$ \\ 
    \hline 
    No Opinion & 6 & 7$\%$ \\ 
    \hline 
    Total & 91 & 100$\%$ \\ 
    \hline 
    \end{tabular}
    \caption{\small Question 3 with some participants responding multiple times (see Figure \ref{fig:Q3} and Table \ref{tab:Q3} for comparison).}
\end{table}

\subsubsection{Testing Quantum Gravity Models with Black Holes }

\begin{table}[H]
    \centering
    \begin{tabular}{|c|c|c|}
    \hline
    \textbf{Answer Option} & Number of Votes & Percentage \\
    \hline
    Black hole shadows are not dark & 8 & 8$\%$ \\ 
    \hline 
    Effective Field Theory corrections to inspiral/ringdown & 30 & 31$\%$ \\ 
    \hline 
    Gravitational Wave Echoes & 13 & 14$\%$ \\ 
    \hline 
    Deviations from Kerr Spacetime Multipoles & 13 & 14$\%$ \\ 
    \hline 
    Other & 24 & 25$\%$ \\ 
    \hline 
    No opinion & 8 & 8$\%$ \\ 
    \hline 
    Total & 96 & 100$\%$ \\ 
    \hline 
    \end{tabular}
    \caption{\small Question 4 with some participants responding multiple times (see Figure \ref{fig:Q4} and Table \ref{tab:Q4} for comparison).}
\end{table}

\subsubsection{Interpretation of Quantum Mechanics}

\begin{table}[H]
    \centering
    \begin{tabular}{|c|c|c|}
    \hline
    \textbf{Answer Option} & Number of Votes & Percentage \\
    \hline
    Copenhagen & 24 & 28$\%$ \\ 
    \hline 
    Many Worlds & 11 & 13$\%$ \\ 
    \hline 
    Qbism & 0 & 0$\%$ \\ 
    \hline 
    Pilot Wave/de Broglie–Bohm & 3 & 3$\%$ \\ 
    \hline 
    Collapse Theories & 6 & 7$\%$ \\ 
    \hline
    Consistent Histories & 4 & 5$\%$ \\ 
    \hline 
    Other & 12 & 14$\%$ \\ 
    \hline 
    No Opinion & 27 & 31$\%$ \\ 
    \hline 
    Total & 87 & $100\%$  \\
    \hline 
    \end{tabular}
    \caption{\small Question 5 with some participants responding multiple times (see Figure \ref{fig:Q5} and Table \ref{tab:Q5} for comparison).}
\end{table}

\subsubsection{Dark Matter}

\begin{table}[H]
    \centering
    \begin{tabular}{|c|c|c|}
    \hline
    \textbf{Answer Option} & Number of Votes & Percentage \\
    \hline
    Light particle dark matter (e.g., axions) & 12 & 13$\%$ \\ 
    \hline 
    Heavy particle dark matter (e.g., WIMPs) & 9 & 10$\%$ \\ 
    \hline 
    Some form of modified gravity & 3 & 3$\%$ \\ 
    \hline 
    Primordial black holes & 14 & 16$\%$ \\ 
    \hline 
    Some hybrid & 23 & 26$\%$ \\ 
    \hline
    Effect of quantum gravity & 6 & 7$\%$ \\ 
    \hline 
    Other & 12 & 13$\%$ \\ 
    \hline 
    No opinion & 11 & 12$\%$ \\ 
    \hline 
    Total & 90 & 100$\%$ \\
    \hline
    \end{tabular}
    \caption{\small Question 7 with some participants responding multiple times (see Figure \ref{fig:Q7} and Table \ref{tab:Q7} for comparison).}
\end{table}

\subsubsection{Early Universe Cosmology}

\begin{table}[H]
    \centering
    \begin{tabular}{|c|c|c|}
    \hline
    \textbf{Answer Option} & Number of Votes & Percentage \\
    \hline
    Cosmic Inflation & 37 & 40$\%$ \\ 
    \hline
    Bouncing or Cyclic Universe & 9 & 10$\%$ \\ 
    \hline 
    Some quantum gravity model & 15 & 16$\%$ \\ 
    \hline
    Breakdown of Lorentz Invariance & 5 & 5$\%$ \\ 
    \hline 
    Not inflation but don’t have an alternative & 10 & 11$\%$ \\ 
    \hline 
    Other & 8 & 9$\%$ \\ 
    \hline 
    No Opinion & 9 & 10$\%$ \\ 
    \hline 
    Total & 93 & 100$\%$ \\
    \hline 
    \end{tabular}
    \caption{\small Question 8 with some participants responding multiple times (see Figure \ref{fig:Q8} and Table \ref{tab:Q8} for comparison).}
\end{table}

\subsubsection{Meaning of the Big Bang}

\begin{table}[H]
    \centering
    \begin{tabular}{|p{0.53\linewidth}| p{0.1\linewidth}|p{0.1\linewidth}|}
    \hline
    \textbf{Answer Option} & Number of Votes & Percentage \\
    \hline
    An absolute beginning of time with a singularity at its start & 10 & 12$\%$ \\ 
    \hline 
    An absolute beginning of time but without a singularity & 5 & 6$\%$ \\ 
    \hline 
    A theory that says the universe evolved from a hot dense state that says nothing about whether there was an absolute beginning of time or not  & 55 & 67$\%$ \\ 
    \hline 
    Other & 8 & $10\%$ \\
    \hline 
    No opinion & 4 & $5\%$ \\
    \hline 
    Total & 82 & $100\%$ \\
    \hline 
    \end{tabular}
    \caption{\small Question 9 with some participants responding multiple times (see Figure \ref{fig:Q9} and Table \ref{tab:Q9} for comparison).}
\end{table}

\subsubsection{Anthropic Coincidences and Fine tuning}

\begin{table}[H]
    \centering
    \begin{tabular}{|p{0.53\linewidth}| p{0.1\linewidth}|p{0.1\linewidth}|}
    \hline
    \textbf{Answer Option} & Number of Votes & Percentage \\
    \hline
    The values of the constants are set by necessity & 8 & 10$\%$ \\ 
    \hline 
    They are brute facts or need no further explanation & 26 & 32$\%$ \\ 
    \hline 
    Explained by anthropic selection in a multiverse & 20 & 25$\%$ \\ 
    \hline 
    Explained by a Darwinian process occurring in the cosmos & 8 & 10$\%$ \\ 
    \hline 
    Explained by an intelligent designer & 2 & 2$\%$ \\ 
    \hline 
    Other & 7 & 9$\%$ \\ 
    \hline 
    No opinion & 10 & 12$\%$ \\ 
    \hline 
    Total & 81 & 100$\%$ \\
    \hline 
    \end{tabular}
    \caption{\small Question 10 with some participants responding multiple times (see Figure \ref{fig:Q10} and Table \ref{tab:Q10} for comparison).}
\end{table}

\subsubsection{Dark Energy}

\begin{table}[H]
    \centering
    \begin{tabular}{|c|c|c|}
    \hline
    \textbf{Answer Option} & Number of Votes & Percentage \\
    \hline
    A cosmological constant & 32 & 37$\%$ \\ 
    \hline 
    A time varying field & 13 & 15$\%$ \\ 
    \hline 
    An effect of quantum gravity & 12 & 14$\%$ \\ 
    \hline 
    Some other modification to gravity & 15 & 17$\%$ \\ 
    \hline 
    Other & 6 & 7$\%$ \\ 
    \hline 
    No option & 8 & 9$\%$ \\ 
    \hline 
    Total & 86 & 100$\%$ \\ 
    \hline 
    \end{tabular}
    \caption{\small Question 11 with some participants responding multiple times (see Figure \ref{fig:Q11} and Table \ref{tab:Q11} for comparison).}
\end{table}

\subsubsection{Hubble Tension}

\begin{table}[H]
    \centering
    \begin{tabular}{|c|c|c|}
    \hline
    \textbf{Answer Option} & Number of Votes & Percentage \\
    \hline
    Systematic Error in Supernova Data & 27 & 35$\%$ \\ 
    \hline 
    Systematic Error in CMB/Baryonic Acoustic Oscillation Data & 4 & 5$\%$ \\ 
    \hline 
    Early Dark Energy & 5 & 6$\%$ \\ 
    \hline 
    Other New Particles or Fields & 1 & 1$\%$ \\ 
    \hline 
    Modified or Quantum Fields & 12 & 16$\%$ \\ 
    \hline 
    Other & 10 & 13$\%$ \\ 
    \hline 
    No opinion & 18 & 23$\%$ \\ 
    \hline 
    Total & 77 & $100\%$ \\
    \hline 
    \end{tabular}
    \caption{\small Question 12 with some participants responding multiple times (see Figure \ref{fig:Q12} and Table \ref{tab:Q12} for comparison).}
\end{table}

\newpage 

\bibliographystyle{plainnat}
\bibliography{bibliography-biblatex}

\end{document}